\documentclass{birkjour}
 \newtheorem{thm}{Theorem}[section]

 \theoremstyle{definition}
 
 \theoremstyle{remark}
 \newtheorem{rem}[thm]{Remark}
 
 \numberwithin{equation}{section}
\usepackage{url}
\usepackage{amsmath,amssymb}
\usepackage{amsfonts}
\usepackage{graphicx} 
\usepackage{subfig} %
\usepackage{caption}

\begin{document}

\title[Logistic growth model and modeling of factors for community case transmission.]
  {Logistic growth model and modeling of factors for COVID-19 community case transmission in Senegal}
\author[Diouf M.]{Massamba Diouf}
\address{Faculty of Medicine, Pharmacy and dentistry, public health department
 \br
University of Cheikh Anta Diop, Dakar, Senegal}
\email{dioufmass78@yahoo.fr, massamba.diouf@ucad.edu.sn}

\author[Ndiaye B.M.]{Babacar Mbaye Ndiaye}
\address{Laboratory of Mathematics of Decision and Numerical Analysis\br
University of Cheikh Anta Diop\br
BP 45087, 10700. Dakar, Senegal}
\email{babacarm.ndiaye@ucad.edu.sn}

\thanks{This work was completed with the support of the NLAGA project}


\date{November 04, 2020}
\begin{abstract}
In this article, we analyze the spread of cases resulting from community transmission of COVID-19 in Senegal in order to identify statistical associations. The identification and knowledge of the factors associated with this community transmission can be a decision support tool to limit the spread of the disease. We estimate parameters and evaluate the growth factor, community rate, weekly increase and daily difference, and make forecasting to help on how to find concrete actions to control the situation.
\end{abstract}

\maketitle
\section{Introduction}
\noindent  In December 2019, the outbreak of COVID-19 associated with a new coronavirus was a significant event. The ensuing spread prompted the World Health Organization to declare a pandemic on March 11, 2020, after it was considered a health emergency of global concern \cite{who1}.
The whole world is thus facing an unprecedented health crisis from the coronavirus with the number of people affected which is gradually increasing with more than 25,000,000 confirmed cases for nearly 800,000 deaths at the end of August 2020 \cite{who2}.\\
Less affected by the pandemic compared to other continents, Africa is still resisting with a COVID-19 death toll reaching nearly 30,000 deaths while confirmed cases have exceeded 1,130,000 (August 31, 2020) according to the African control center and disease prevention  \cite{cdc}.\\
This progression of the disease in this zone raises concerns as well among the governors as among the health experts with regard to the living conditions of the communities, the insufficient resources to face this type of threat and the noted failures. in some health systems in Africa. 
In Senegal, at the start of the epidemic, the government declared a state of emergency with several health, economic and security measures taken to reduce the spread of COVID-19. Then in the face of socio-economic stress, an easing followed by the lifting of restrictive measures seems to give a boost to the disease with the meteoric increase in incident cases. 
From March to April, the number of cases was 933. In the period from May-June, people sick with the coronavirus increased by 5,860. From July to August, the number of cases in Senegal increased by 6,818. This progression affecting almost all the health districts of all the regions of the country results for the most part from the resurgence of cases resulting from community transmission. These community cases are characterized by their difficulty in tracing to identify the source of contamination, but also by the limited possibilities for identifying and isolating the various associated contacts. During the three periods described above, the number of community cases among people positive for the coronavirus in Senegal was 109 for the first two-month period, 709 for the second two-month period and for the third two-month period July-August 2384. This development reflects an epidemiological situation including the magnitude and seriousness with regard to the consequences related to lethality and other impacts of all kinds, remain worrying. The determinants and/or variables linked to this form of transmission are not sufficiently documented in Senegal. Therefore, the identification and knowledge of the factors associated with this community transmission can constitute a decision support tool to limit the spread of the disease, on the one hand through targeted actions, to control the epidemic and its externalities on the other hand. The objective of this work is to model the variables potentially involved in the spread of cases resulting from community transmission of COVID-19 in Senegal in order to identify statistical associations.\\
In our previous articles \cite{ndiaye1,ndiaye2,ndiaye3,ndiaye4,ndiaye5}, we try to forecast the pandemic for Senegal using the SIR, stochastic SIR and machine learning. 
Here, we give forecasting pandemic size of community cases for Senegal and daily predictions using the logistic model. We assume that the model is a reasonable description of the epidemic. Full daily reports are available in \cite{msas}.
\noindent First, we collect the pandemic data from \cite{msas}, from March 03, 2020 to September 20, 2020. Then, we propose a logistic growth model for forecasting.
Variables such as community rate, growth factor, daily difference (daily variations) and weekly increase (number of cumulative cases per week) were defined to better understand the dynamics of the development of community cases in Senegal.\\
\noindent The article is organized as follows. In section \ref{analysis}, we present some data analysis followed by the Logistic growth model for forecasting in section \ref{logistic}. Finally, in section \ref{ccl}, we present conclusions and perspectives.

\section{Data analysis}\label{analysis}
\begin{figure}[h!]
	\centering
	\includegraphics[width=.9\linewidth]{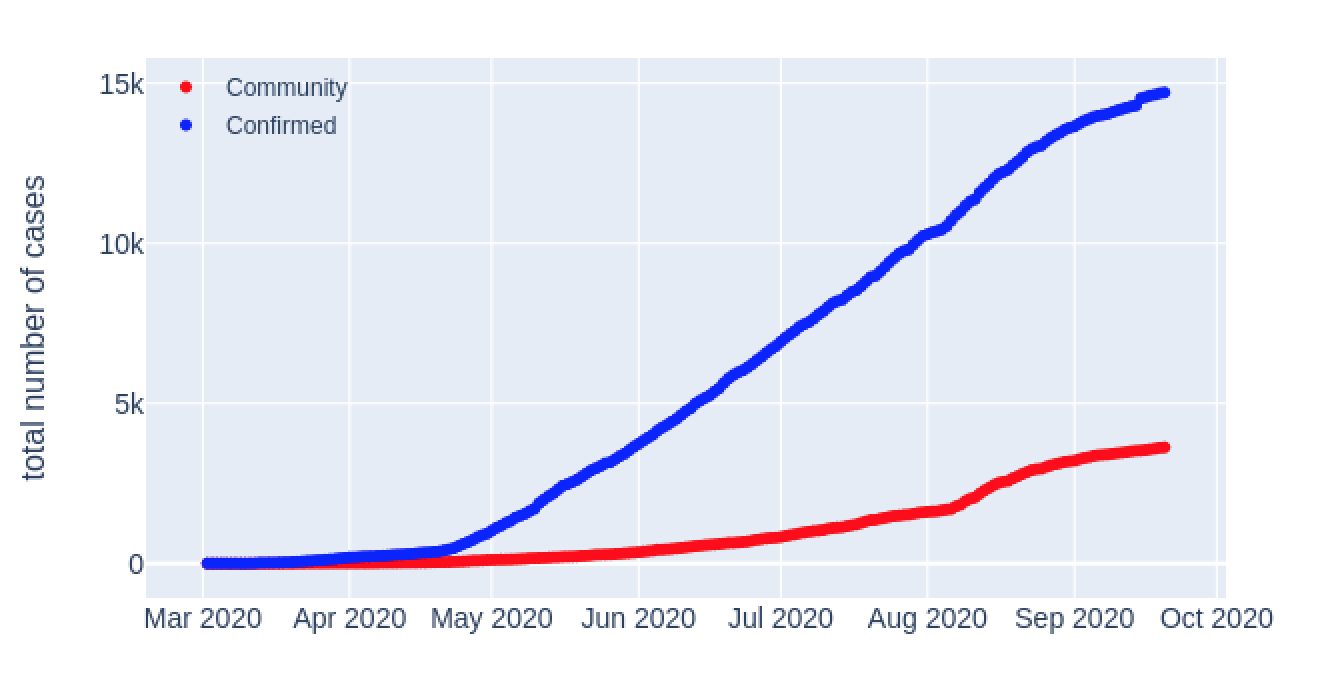}
		\par\vspace{-0.5cm}
	\caption{Senegal: confirmed and community cases}\label{ccom_sn}
\end{figure}

The simulations are carried out from  data in \cite{msas}, from March 02, 2020, to September 20, 2020. 
The numerical tests were performed by using the Python with Panda library 
\cite{python}, and were executed on a computer with the following characteristics: intel(R) Core-i7 CPU 2.60GHz, 24.0Gb of RAM, under UNIX system.\\
According to daily reports, we first analyze and make some data preprocessing before simulations. The cumulative numbers of confirmed and community cases are illustrated in Figure \ref{ccom_sn}, which shows a progression of community cases as a function of the total number of confirmed cases. 
\begin{figure}[h!]
	\centering
	\includegraphics[width=.9\linewidth]{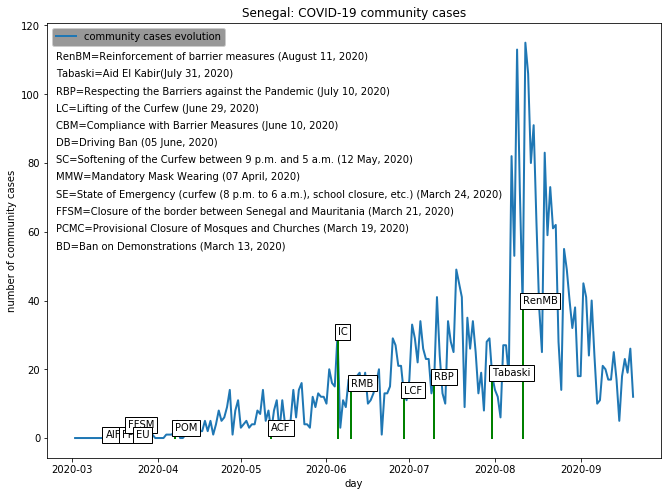}
		\par\vspace{-0.5cm}
	\caption{Senegal:  community cases per day}\label{ccom_C_dates_sn}
\end{figure}
It seems to mean that the increase in confirmed cases correlates with the increase in community cases.\\
Figure \ref{ccom_C_dates_sn} shows the variation in the number of community cases with the various institutional measures taken since the declaration of the index community case on March 21, 2020; as well as the religious events that occurred during this period.\\
The restriction of barrier measures has helped stabilize community cases (RBP, July 10, 2020 from Figure \ref{ccom_C_dates_sn}), with a slight increase followed by a decrease.\\
The Aid Al Kabir (Tabaski, July 31, 2020) intensively increases the cases, because during this period people mix from house to house.
This increase in cases has prompted the government to strengthen barrier measures (RenBM, August 31, 2020). This increase in cases has prompted the government to strengthen barrier measures (RenBM, August 31, 2020). That made it possible to reduce the transmission.\\
In managing the pandemic, controlling movement is a sure way to limit the spread. This is why slogans urging everyone to stay at home were widely shared on social media to persuade more.

\subsection{Community rate}
\begin{figure}[h!]
	\centering
	\includegraphics[width=.9\linewidth]{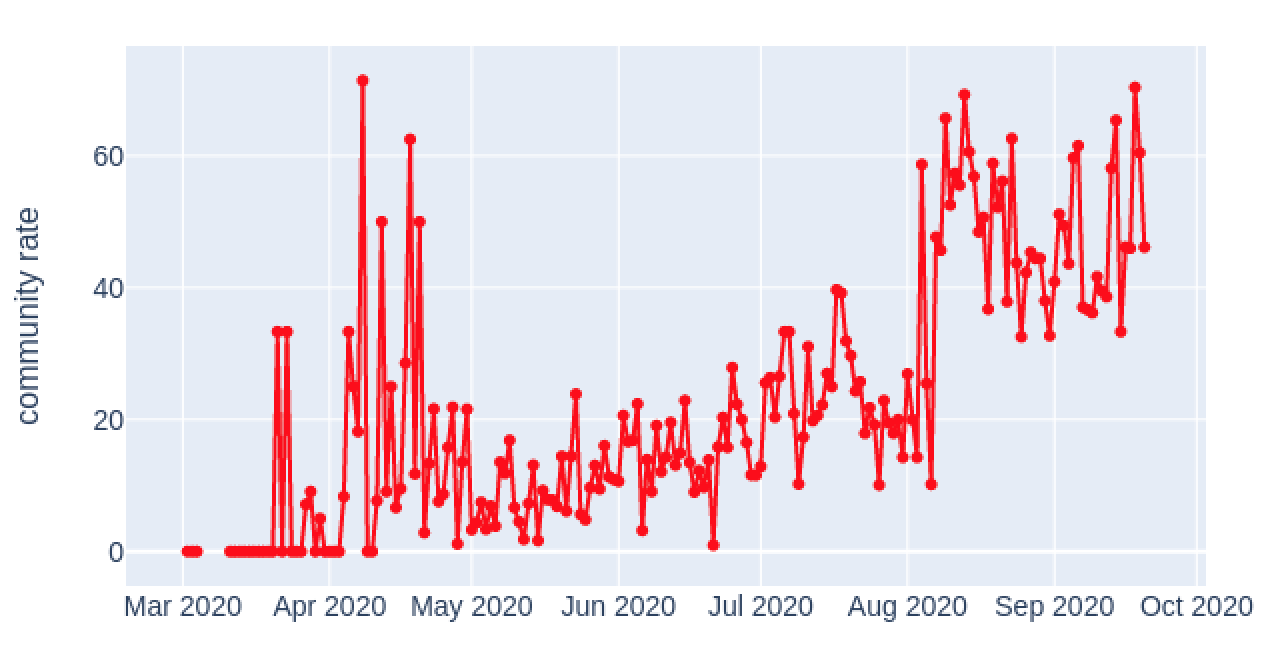}
		\par\vspace{-0.5cm}
	\caption{Senegal: Community rate per day}\label{rate_sn}
\end{figure}

First, let's define the community rate. 
$$
\mbox{Community rate} = \frac{\mbox{number of community cases}}{\mbox{number of confirmed cases}} \times 100
$$
The community rate (per day) is given by Figure \ref{rate_sn}, and it's average and median values in Table \ref{sn_avermedvalues}. 
\noindent The analysis of Figure \ref{rate_sn} shows a variable evolution of the rate with extreme values in April and August-September.
\begin{table}[h!] 
\begin{center}
\begin{tabular}{|l|c|} 
 \hline
 {\bf average/median} 	& {\bf values}  \\ \hline
average community rate &   22.562284  \\\hline
median community rate &   17.931034 \\\hline
\end{tabular}
\end{center}
\caption{Average and median values}\label{sn_avermedvalues}
\end{table}

\subsection{Growth factor}
\begin{figure}[h!]
	\centering
	\includegraphics[width=.9\linewidth]{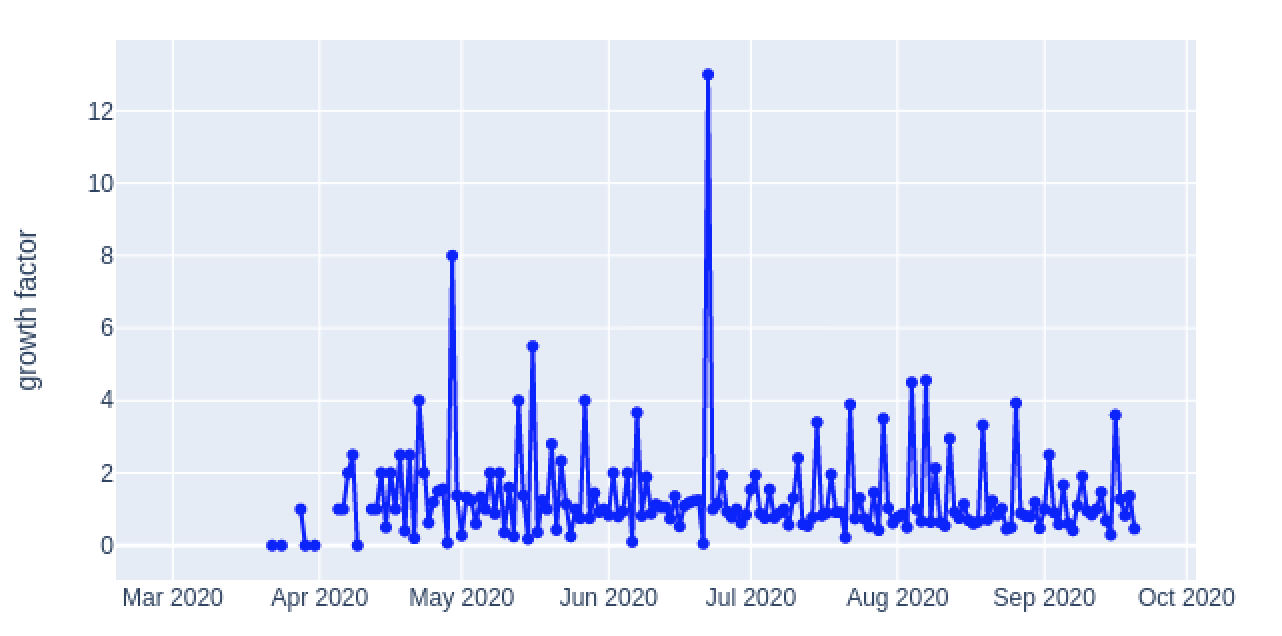}
		\par\vspace{-0.5cm}
	\caption{Senegal: Growth factor community cases}\label{growthfac_sn}
\end{figure}

The growth factor is the factor by which a quantity multiplies itself over time. The formula used is: $$
 \qquad \frac{\mbox{Every day's new community cases}}{\mbox{new community cases on the previous da}y}
$$
\begin{rem}
\begin{itemize}
\item A growth factor constant at 1 indicates there is no change in any kind of cases.
\item A growth factor above 1 indicates an increase in corresponding cases.
\item A growth factor above 1 but trending downward is a positive sign, whereas a growth factor constantly above 1 is the sign of exponential growth.
\end{itemize}
\end{rem}
\noindent The growth factor for community cases is given by Figure \ref{growthfac_sn}, and it's  average and median growth factors in Table \ref{sn_avermedincrease}. The two largest growth peaks (8 and 13) are reached in May and July.
\begin{table}[h!] 
\begin{center}
\begin{tabular}{|l|c|} 
 \hline
 {\bf average/median growth factor} 	& {\bf values}  \\ \hline
average growth factor of number of community cases & inf  \\ \hline
median growth factor of number of community cases &  1.0 \\ \hline 
\end{tabular}
\end{center}
\caption{Average and median growth factors of community cases }\label{sn_avermedincrease}
\end{table}

\subsection{Daily difference}
\begin{figure}[h!]
	\centering
	\includegraphics[width=.9\linewidth]{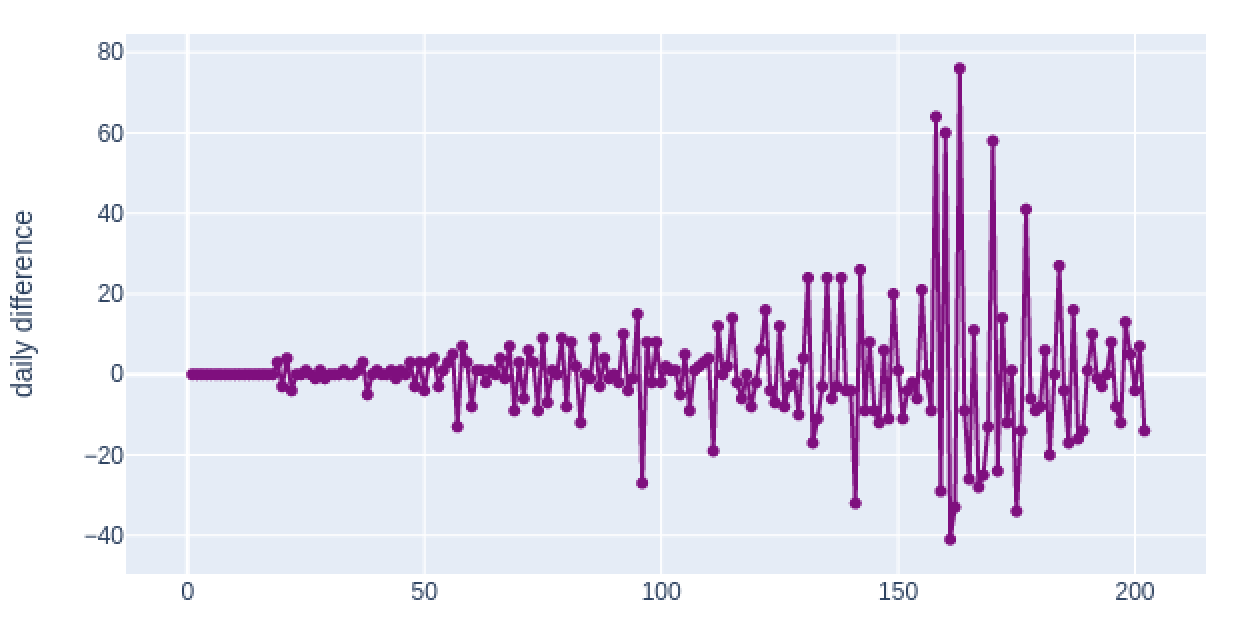}
		\par\vspace{-0.5cm}
	\caption{Senegal: Daily difference}\label{dailydif_sn}
\end{figure}

\noindent The Daily difference for community cases is given by Figure \ref{dailydif_sn}. 
\noindent The greatest variations are obtained between the end of July and the beginning of August (period of after Aid El Kabir).

\subsection{Weekly increase}
\begin{figure}[h!]
	\centering
	\includegraphics[width=.9\linewidth]{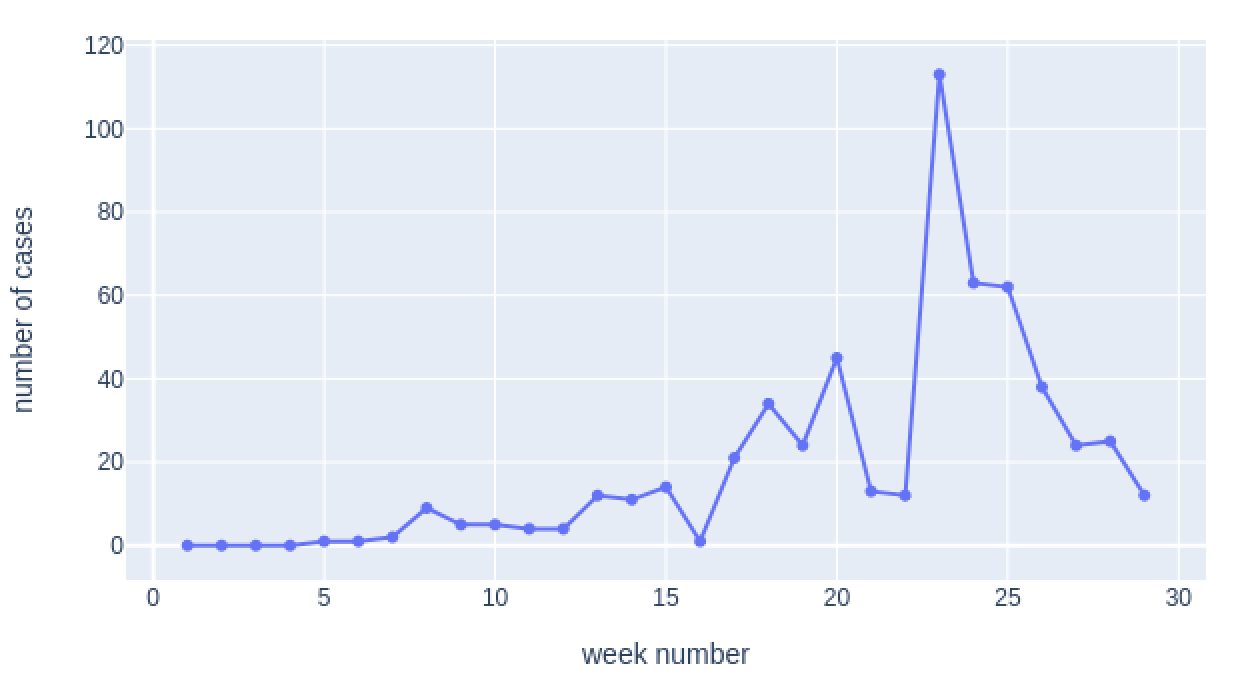}
		\par\vspace{-0.5cm}
	\caption{Weekly growth of community cases}\label{weeksC_sn}
\end{figure}

\begin{figure}[h!]
	\centering
	\includegraphics[width=.8\linewidth]{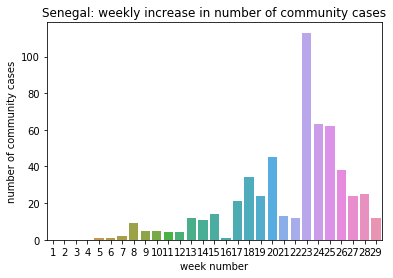}
	\par\vspace{-0.5cm}
	\caption{Weekly increase in number of community cases}\label{weeksH_sn}
\end{figure}

The curve and the histogram for the weekly increase in number of community cases are illustrated in Figures \ref{weeksC_sn} and \ref{weeksH_sn}, respectively.
\newpage
\noindent We see that week 23 was fatal for community cases, which corresponds to the period after Aid El Kabir.

\section{Logistic grow model for community cases}\label{logistic}
In this section,  we a perform forecast with the logistics model  (\cite{wang,chowell,brauer}) to predict the final size of coronavirus epidemy, for community cases in Senegal. \\
The logistic model is a data-driven phenomenological model. Thus its predictions are as good as useful data are and as good as it can mimics epidemy dynamics. 
It's in a way mimic both epidemy spreading and its control i.e. prevention measures. When daily predictions of epidemic size begin to converge, we can say that epidemy is in control.

\subsection{The model}
In mathematical epidemiology, the epidemic dynamics can be described by the following variant of logistic growth model:
\begin{equation}\label{logmodel}
\frac{dC}{dt}=rC\Big(1-\frac{C}{K}\Big)
\end{equation}
where $C$ is an accumulated number of community cases, $r>0$ infection rate, and $K>0$ is the final epidemic size (of community cases). If $C(0)=C_0>0$ is the initial number of cases then the solution of (\ref{logmodel}) is:
\begin{equation}\label{logmodelsol}
C = \frac{K}{1+Ae^{-rt}}
\end{equation}
where $A = \frac{K-C_0}{C_0}$. \\
When $t\ll  1 $, assuming $K\gg C_0$, and therefore $A \gg 1$. 
We have the natural growth:
$$
C = \frac{Ke^{rt}}{e^{rt}+A} = \Big(\frac{C_0}{1-C_0/K}\Big) \Big(\frac{e^{rt}}{1+e^{rt}/A}\Big) \approx C_0e^{rt}
$$
When $t \longrightarrow \infty  $ the number of community cases follows the Weibull function: 
$$
C = K \big( 1-Ae^{-rt} +... \big) \approx K   \Big[1-e^{-r(t-t_0)} \Big]
$$
The growth rate $\frac{dC}{dt}$ reaches its maximum when $\frac{d^2C}{dt^2}=0$.  From this condition, we obtain that the growth rate peak occurs in time: 
$$
t_p = \frac{\mbox{ln} A}{r}
$$
At this time the number of cases is
$
C_p = \frac{K}{2}
$
and the growth rate is
$$
\Big(\frac{dC}{dt}\Big)_p  = \frac{rK}{4}
$$
To answer the question about doubling time $\Delta t$, i.e., the time takes to double the number of cases we solve $C(t+\Delta t)=2C(t)$ for $\Delta t$ . Result is
$$
\Delta t = \frac{\mbox{ln} 2}{r} - \frac{1}{r}\mbox{ln} \Big(\frac{1}{A} - e^{-rt} \Big) - t 
$$
The first term represents initial exponential growth, then $\Delta t$  increases with $t$. When $t \longrightarrow t_p = \frac{\mbox{ln} A}{r}$, i.e., when  $C \longrightarrow \frac{K}{2} $, then $\Delta t \longrightarrow \infty$. 
For $C\geq \frac{K}{2}$ doubling time lost its meaning. 
For more details for the logistic growth model, see \cite{wang,chowell,brauer}.

\subsection{Numerical analysis}
\begin{figure}[h!]
	\centering
\includegraphics[width=.9\linewidth]{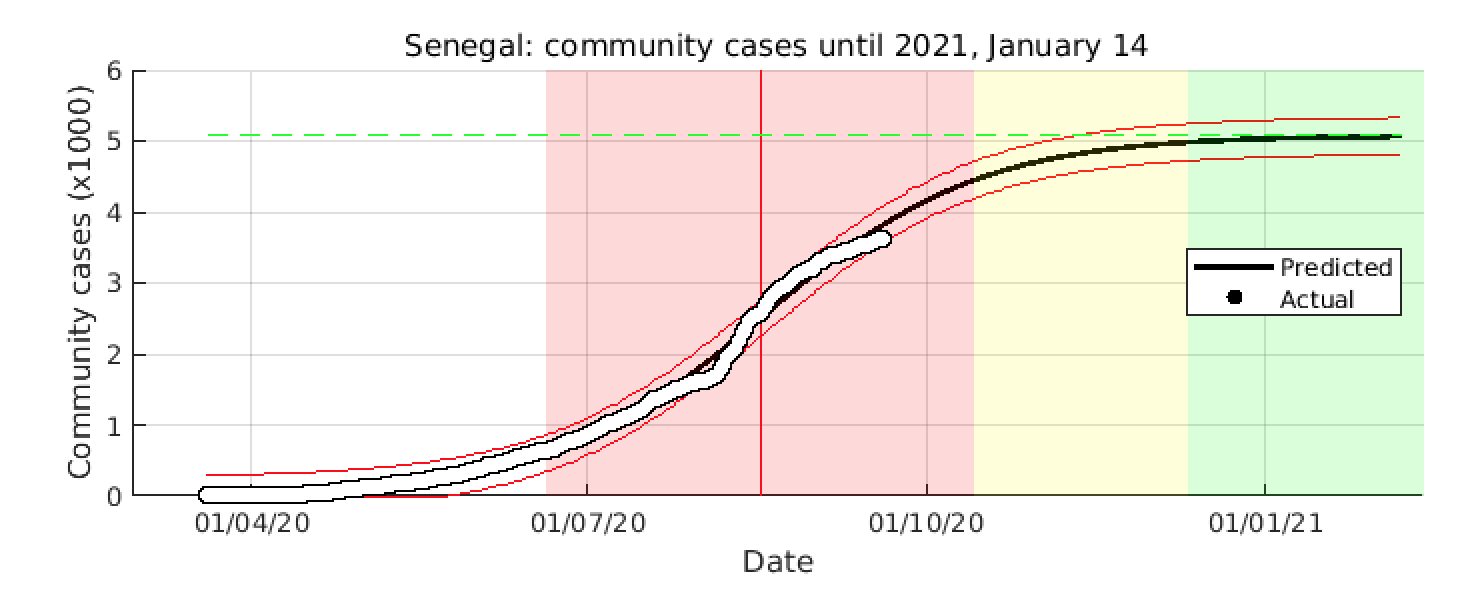}
	\caption{Senegal: logistic growth model forecasting of community cases}\label{forecast_sn}
\end{figure}

\begin{figure}[h!]
	\centering
\includegraphics[width=.9\linewidth]{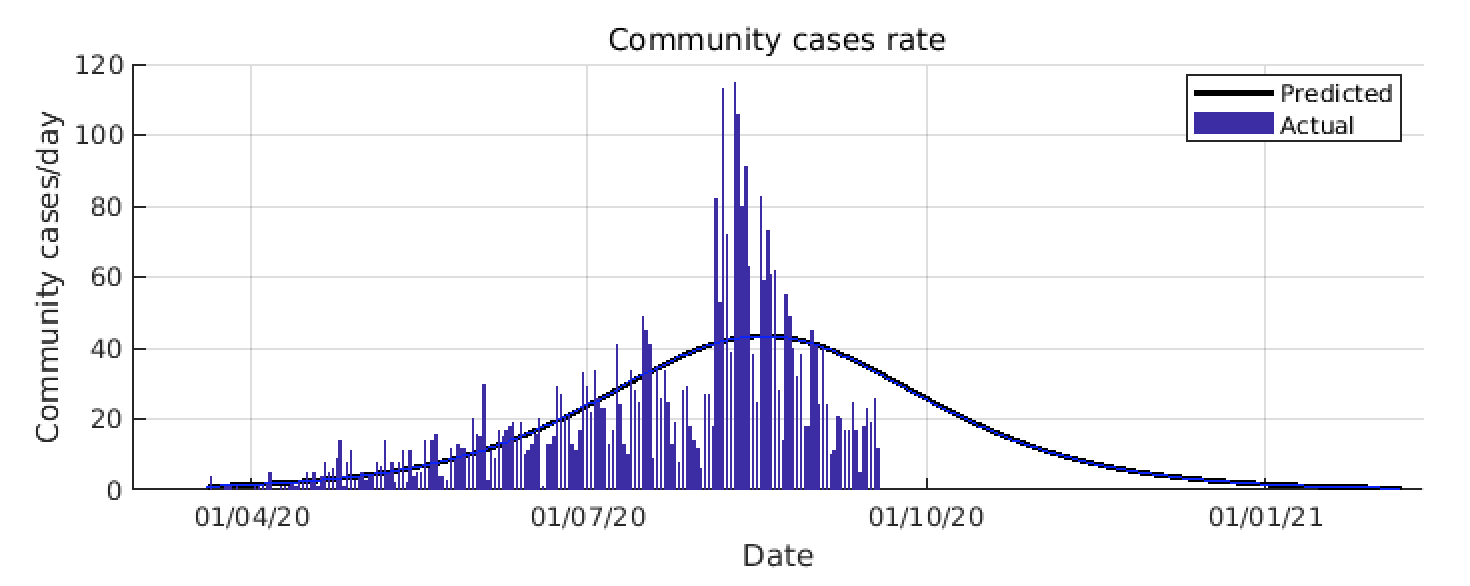}
	\caption{Senegal: forecasted final size of the community cases}\label{fig_finalsize}
\end{figure}

In Senegal, the first case of coronavirus appeared on March 2, 2020. But the first community case was detected on March 21, 2020. 
First, we collect the pandemic data from \cite{msas}, from March 21, 2020, to September 20, 2020. Then, we perform a forecast with the logistics model to predict the final size of coronavirus epidemy, for community cases. \\
The size will be approximately 5100 $\mp$ 130 cases (Table \ref{tab_finalsize}). The estimated coefficients are given in Table \ref{tab_finalsize}, and that the peak of the epidemic was on August 12, 2020. 
It seems that the epidemic in Senegal is in the ending stage (see Figures \ref{forecast_sn} and \ref{fig_finalsize}).\\
The short-term forecasting is given in Table \ref{tab_shorterm} where we see that the discrepancy of actual and forecasted number of cases is within 3\%. However, actual and predicted daily new cases are scattered and vary between 26\% to 600\%. \\
On September 25, 2020, the actual number of cases was 3697, and the daily number of cases was 15. Prediction in Table \ref{tab_shorterm} is cumulative 3993 cases and 29 daily cases. The errors are 48.276\% and 7.413\%, respectively.

\begin{table}[h!] 
\begin{center}
\begin{tabular}{|l|c|c|c|c|} 
\hline
 	&  {\bf Estimate} &   {\bf SE}    &  {\bf tStat}    & {\bf pValue } \\ 
 \hline
 K &      5079.3 &       129.32 &   39.278   &  1.6383e-90  \\\hline
    r    &0.034157  &  0.00071248   & 47.941  &  8.3375e-105 \\ \hline
    A     & 173.03 &       10.442   & 16.571 &    3.9785e-38 \\	\hline
\end{tabular}
\end{center}
\caption{Senegal: estimated logistic model parameters}\label{tab_finalsize}
\end{table}
\par\vspace{-0.5cm}
\begin{tabular}{l} 
{\tt 
Number of observations: 184, Error degrees of freedom: 181}\\
{\tt Root Mean Squared Error: 86.4}\\
{\tt R-Squared: 0.995,  Adjusted R-Squared 0.995}\\
{\tt F-statistic vs. zero model: 2.17e+04, p-value = 3.41e-231}
\end{tabular}
\begin{table}[h!] 
\begin{center}
\begin{tabular}{|c|c|c|c|c|c|c|c|} 
\hline
 {\bf Day}     &    {\bf Date}  &   {\bf Actual}  &  {\bf Predict}  & {\bf Error \% } & {\bf Daily } & {\bf Daily}     & {\bf Error \%} \\ 
      &     &     &    &    & {\bf actual} & {\bf predicted}    & \\ 
 \hline
 179  &14-Sep-2020 &      3527    &   3638  &     3.15   &       5    &     35   &   600.00\\ \hline
 180  &15-Sep-2020   &    3545    &   3673  &     3.61    &     18     &    35    &   94.44\\ \hline
 181  &16-Sep-2020  &     3568    &   3708   &    3.92     &    23     &    35    &   52.17\\ \hline
 182  &17-Sep-2020   &    3587   &    3742    &   4.32    &     19    &     34   &    78.95\\ \hline
 183  &18-Sep-2020    &   3613  &     3775     &  4.48     &    26    &     33    &   26.92\\ \hline
 184  &19-Sep-2020    &   3625  &     3808    &   5.05    &     12     &    33   &   175.00\\ \hline
 185  &20-Sep-2020     &     -     &  3840    &      -    &      -        & 32  &     \\ \hline
 186  &21-Sep-2020     &     -     &  3872       &   -      &    -        & 32  &    \\ \hline
 187  &22-Sep-2020     &     -      & 3903      &    -      &    -      &   31  &\\ \hline
 188  &23-Sep-2020      &    -      & 3934      &    -      &    -       &  31  &\\ \hline
 189  &24-Sep-2020      &    -     &  3964       &   -      &    -      &   30  &\\ \hline
 190  &25-Sep-2020       &   -      & 3993     &     -       &   -       &  29  &\\ \hline
\end{tabular}
\end{center}
\caption{Senegal: short-term forecasting}\label{tab_shorterm}
\end{table}

\section{Conclusion and Perspectives}\label{ccl}
These extensive statistical analyzes allow us to make predictions about the future of the pandemic in Senegal. In the first phase of the response (at the national level), the emphasis was more on government actions, the involvement of health professionals, and uncompromising communication. Then, with the increase in cases of so-called community transmission, the epidemic entered its active phase with an increasing incidence in society. This worrying situation has led to the strengthening of various measures: the state of emergency with a curfew, the prohibition or restriction of interurban travel, gatherings but above all barrier gestures such as wearing a mask. This context of restriction, like everywhere in the world, has caused a gloomy economic environment in Senegal forcing the State to readjust measures. An easing of restrictions is announced first and then, a little later, the lifting of institutional measures, without first giving the necessary guarantees and support for the response strategy at the global level. Decisions that seemed to help boost the spread of the pandemic nationally with areas, initially free from the disease, which recorded their first cases of COVID-19.\\
Despite these considerations, religious events like the Eid feast have taken place during this critical period. The festival of Eid or sacrifice of the sheep, in Senegalese sociology is a symbol and a pillar of stability and solidarity. The individuals preferentially celebrate it in a family or in the community. It is a time of exchanges of civilities, prayers in communion and sharing meals. This could facilitate the breaking of the barrier measures against the spread of COVID-19. Moreover, one to two weeks after the July 31 celebration of this holiday in Senegal, the rate of contamination of the disease has almost tripled with all the affected regions and the resurgence of cases resulting from community transmission.
This situation further prompted the state to make the general wearing of masks compulsory in public places, commerce, transport, schools and universities. Wearing a mask remains a beneficial measure for many countries and reduces contamination by limiting the projection of postilions (especially in public). This fact seems to contribute to the control of the disease in Senegal which, since the beginning of September, has been in a downward dynamic. \\
According to an optimistic estimate, the pandemic in Senegal will end soon, while for most countries of the world, the blow of the anti-pandemic will be no later than the end of the year.


%

\end{document}